\documentclass[conference]{IEEEtran}
 \IEEEoverridecommandlockouts
\usepackage{cite}
\usepackage{amsmath,amssymb,amsfonts}
\usepackage{graphicx}
\usepackage{textcomp}
\usepackage{xcolor}
\usepackage{comment}
\usepackage{optidef}
\usepackage[T1]{fontenc}
\usepackage{mathtools} 
\usepackage{cuted}
\usepackage{bbm}
\usepackage{dsfont}
\usepackage{mathtools}

\begin{document}
\title{Deep Learning-Based Active User Detection for Grant-free SCMA Systems
\thanks{This work was supported by the Academy of Finland 6Genesis Flagship (grant no. 318927).}}
 \author{\IEEEauthorblockN{Thushan Sivalingam, Samad Ali, Nurul Huda Mahmood, Nandana Rajatheva, and Matti Latva-Aho}
 \IEEEauthorblockA{Centre for Wireless Communications, University of Oulu, Oulu, Finland}
\{thushan.sivalingam, samad.ali, nurulhuda.mahmood, nandana.rajatheva, matti.latva-aho\}@oulu.fi, 
 }
\maketitle
\begin{abstract}
Grant-free random access and uplink non-orthogonal multiple access (NOMA) have been introduced to reduce transmission latency and signaling overhead in massive machine-type communication (mMTC). In this paper, we propose two novel group-based deep neural network active user detection (AUD) schemes for the grant-free sparse code multiple access (SCMA) system in mMTC uplink framework. The proposed AUD schemes learn the nonlinear mapping, i.e., multi-dimensional codebook structure and the channel characteristic. This is accomplished through the received signal which incorporates the sparse structure of device activity with the training dataset. Moreover, the offline pre-trained model is able to detect the active devices without any channel state information and prior knowledge of the device sparsity level. Simulation results show that with several active devices, the proposed schemes obtain more than twice the probability of detection compared to the conventional AUD schemes over the signal to noise ratio range of interest.
\end{abstract}
\vspace{3mm}
\begin{IEEEkeywords}
Grant-free, active user detection, deep neural network, non-orthogonal multiple access, SCMA.
\end{IEEEkeywords}

\section{Introduction}
Currently, there is a significant interest in machine-type communication (MTC) due to the evolution of Internet-of-Things (IoT) applications for the extended human requirements towards digitalization~\cite{6GFlagship_WP}. This includes eHealth, intelligent transport, smart wearables, smart environments, to name a few. Massive MTC (mMTC) focuses on supporting a massive number of low-power and low-complexity devices that sporadically transmit short data packets~\cite{8663999} with low transmission rates. mMTC traffic is usually sporadic with a small payload. This makes the conventional scheduling-based multiple access scheme in which the base station (BS) allocates orthogonal time/frequency resources to each device through a time-consuming handshaking process, to be not as efficient because of the heavy signaling overhead~\cite{6525600}, higher delay, and collisions in random access process~\cite{8663999}.

To overcome these limitations, grant-free non-orthogonal multiple access (NOMA) schemes have been proposed as a promising solution~\cite{7263349}. To minimize the signaling overhead and increase resource utilization, grant-free access allows the machine-type devices (MTD) to transmit data to the BS without acquiring a grant beforehand. NOMA techniques have been extensively investigated to potentially support a massive number of MTDs by sharing a limited amount of resources in a non-orthogonal fashion~\cite{7263349}. In this approach, there is inter-user interference because of the orthogonality violation. NOMA applies device-specific non-orthogonality sequences to control the inter-user interference~\cite{7263349}. From this perspective, the sparse coded multiple access (SCMA)\footnote{Numerous code-domain (CD)-NOMA schemes have been proposed in recent years. Low-density signatures (LDS)-NOMA performs MUD based on the message passing algorithm (MPA) by exploiting the sparsity of LDS. SCMA extends this concept by introducing sparse codebooks for each user, where each codeword represents a different member of the input constellation leading to an improved error performance~\cite{liu2020sparse}.} is designed as a CD-NOMA technique~\cite{6966170}. Since the BS is unaware of the MTDs transmitting the data, there is a requirement to distinguish the active devices among all potential devices. This procedure is generally called active user detection (AUD).

The transmit vector is defined as a vector representing the activities of all the devices, with $1$ and $0$ corresponding to active and non-active devices, respectively. Due to the sporadic nature of mMTC traffic, only a few devices that are active among a massive number of devices tend to transmit the information simultaneously. Thus, the transmit vector can be readily modeled as a sparse vector. By exploiting the sparsity of this transmit vector, the AUD problem can be formulated as a sparse signal recovery problem. 

Compressed sensing (CS) method has been used frequently to solve the AUD problem~\cite{7842611,7976275,7551125}. The authors of~\cite{7842611} have proposed orthogonal matching pursuit (OMP) and block orthogonal matching pursuit (BOMP) based algorithms to solve the AUD problem. To improve the multi-user detection (MUD) performance, the authors in~\cite{7976275} propose a prior-information-aided adaptive subspace pursuit (PIA-ASP) algorithm to exploit the intrinsically temporal correlation of active user support sets in several continuous-time slots. Meanwhile, the work in~\cite{7551125} presents a dynamic CS-based MUD to realize both user activity and data detection in several continuous-time slots by exploring the temporal correlation of the active user sets. However, all the above mentioned approaches~\cite{7842611,7976275,7551125} assume that the uplink channel state information (CSI) for all the MTDs is perfectly known as a priori to the BS. This is not a practical assumption. Furthermore, the performance of CS-based AUD is poor when the correlation of the system matrix increases and the performance degradation would be high when the sparsity of the input vector increases in mMTC. In addition, these iterative algorithms are formulated and optimized for theoretically guaranteed convergence without considering the time constraint. Taken together, applying iterative algorithms to MTC would increase the communication latency. 

Recent advances in ML research have been able to solve many critical optimization problems in the wireless domain~\cite{6GFlagship_WP2}. There are a few deep learning-based approaches proposed for AUD~\cite{8968401,9149252}. The authors in~\cite{8968401} proposed deep neural network-based AUD for grant-free LDS NOMA schemes, consisting of three parallel receiving units and softmax sparsity estimation. However, using three parallel receivers cause an additional cost in the implementation stage. Furthermore, the proposed sparsity estimation is based on a threshold value to the softmax outputs that perform better only for a particular number of active devices. However, the error is substantial with a different number of active devices. For example, the minimum threshold for two active devices causes the wrong prediction when the actual number of devices is three. Furthermore, the authors in~\cite{9149252} also proposed deep neural network-based AUD and channel estimation for LDS NOMA schemes. It is, therefore, clear that detecting active users without the knowledge of the sparsity variation while avoiding the signaling overhead and maximizing the utilization of the codebook are challenging open problems in grant-free SCMA for mMTC. There is a need for different deep learning approaches for SCMA systems which are more complex than LDS codewords. 

Our main contributions are as follows. We propose a group-based codebook assignment for mMTC that enhances the utilization of the SCMA codebook, and we design a deep neural network-based AUD for a group of MTDs that learns the complicated mapping between the received signal and the transmit vector of the MTDs in a group. Specifically, we jointly detect the number of active MTDs and their identity. Here the neural network has a large number of hidden layers, which can improve the representation quality of the transmit vector. Finally, the proposed solution under the group-based mMTC scenario is evaluated via numerical simulations. In our paper, Section II presents the system model and problem formulation. In Section III, we introduce the proposed deep neural network-based AUD. Numerical results are presented in Section IV. Finally, conclusions are given in Section V.

\subsubsection{Notation}
Lower case letters, boldface lowercase and uppercase letters denote scalars,vectors and matrices, respectively, and calligraphy letters denote sets. The operation $(.)^T$ denotes the transpose. \begin{math} \mathbb{C} \end{math} and \begin{math} \mathbb{R} \end{math} denote the space of complex numbers and real numbers, respectively. $\Re(c)$ and $\Im(c)$ are the real and imaginary part of $c$, respectively. The complex Gaussian distribution with zero mean, variance ${\sigma^2}$ is denoted by \begin{math} \mathcal{CN} (0,{\sigma^2}) \end{math}. The Hadamard (element-wise) product operator is denoted by $\circ$. Finally, the absolute value of the complex number $x$ is denoted by $|x|$ and Euclidean norm of the vector $x$ is denoted by ${\parallel x \parallel}$.

\section{System Model and Problem Formulation}
\subsection{System Model}
Consider an uplink grant-free NOMA of a mMTC scenario with a single BS serving a set $\mathcal{N}$ of $N$ MTDs. The BS and MTDs are each equipped with a single antenna. We consider the overloaded mMTC scenario, where the number of devices are larger than the number resources $L$ \begin{math} (L < N) \end{math}. We assume mMTC devices are active sporadically, and $m$ devices are active in each frame, and devices are synchronized in time. Each device can transmit the pilot and data symbols without scheduling and the BS needs to identify the active devices in each time.
\begin{figure}[t]
    \center
    \includegraphics[width=\linewidth]{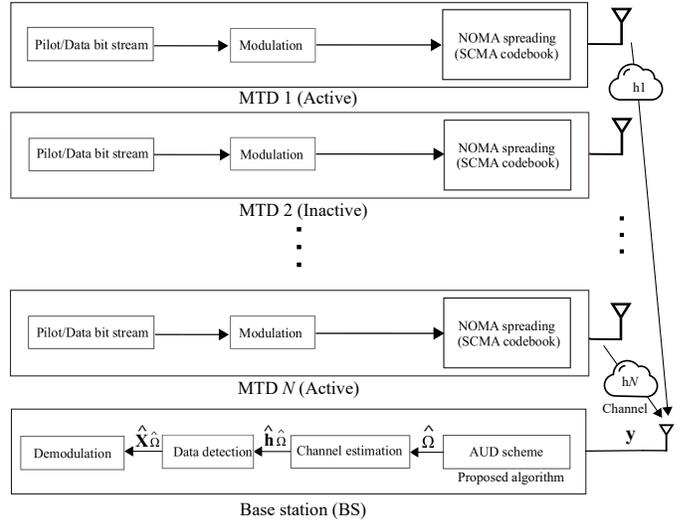}
    \caption{Block diagram of the deep learning based AUD scheme.}
    \label{fig:Proposed_AUD}
\end{figure}

All active devices transmit the information after the spreading with the device-specific non-orthogonal codebook~\cite{9097306} shown in Fig.~\ref{fig:Proposed_AUD}. In this work, we consider the SCMA codebook where $m$ MTDs superimpose their transmissions over $L$ resource elements (REs) according to the sparse mother constellation (MC). The codebook has $L$ dimensional codewords with only $J$ nonzero element assigned to each MTD. Furthermore, the SCMA codebook has a large number of zero elements. Therefore each symbol is spread into only small number of resources, which introduces the necessary sparsity to ensure the decoding quality and reduces the inter-device interference. The relation between the MTDs and REs can be described by a factor graph, where the connection link between the MTD $n$ and RE $l$ indicates the $l$th component of the codeword $n$ is a non-zero element. 

\subsection{Problem Formulation}
We define the binary MTD activity indicator $a_n$ for the $n$th device as
\begin{equation}
 {a_n} = 
\begin{cases}
1, \ $if the $n$th MTD is active,$\\
0, \ $otherwise.$
\end{cases}
\end{equation}
The activity of each device is independent to each other. The received signal at the BS is given by 
\begin{equation}
\textbf{y} = \sum_{n=1}^{N}{a_n}{\textbf{c}_n}{{h}_n}{x}_n + \textbf{w},
\end{equation}
where \begin{math} {\textbf{c}_n} \in \mathbb{C}^L \end{math} is the SCMA codeword vector of the $n$-th device, $h_n$ is the complex uplink channel coefficient from $n$-th device to the BS, $x_n$ is the transmit symbol of the $n$-th device, and \begin{math} \textbf{w} \sim \mathcal{CN} (0,{\sigma_w^2}\textbf{I})\end{math} is the complex-Gaussian noise vector. The grant-free multiple access protocol consists of two phases in each time slot. In first step, each active device transmits the pilot symbol $x_{p,n}$ to the BS. The BS jointly detects the active devices and corresponding channels. In the second step, the active devices transmit $K$ data symbols $x_{d,n}^{[1]}, \dots, x_{d,n}^{[k]}$  to the BS. Finally, BS decodes the data symbol using the obtained device activity and channels. The pilot measurement vector \begin{math} \textbf{y}_{p} \end{math} is given by
\begin{equation}
\textbf{y}_{p} = \sum_{n=1}^{N}{a_n}{\textbf{c}_n}{{h}_n}{x}_{p,n} + \textbf{w}_p.
\end{equation}
Let us define $\boldsymbol{\phi}_n$ = $\textbf{c}_n$${x}_{p,n}$, and $\boldsymbol{\Phi}$ = [{$\boldsymbol{\phi}_1$}, \dots, {$\boldsymbol{\phi}_N$}], we can rewrite (3) as
\begin{equation}
\begin{split}
    \textbf{y}_{p} &  = \sum_{n=1}^{N}{\boldsymbol{\phi}_n}{a_n}{{h}_n} + \textbf{w}_p, \\
    & = \Phi(\textbf{a} \circ \textbf{h}) + \textbf{w}_p,
\end{split}
\end{equation}
where \textbf{a} = $[a_1, \dots, a_N]^T$ is the activity vector, and \textbf{h} = $[{h}_1, \dots, {h}_N]^T$ is the channel vector. Furthermore, we define the vector \textbf{g} = $(\textbf{a} \circ \textbf{h})$. We can rewrite (4) as
\begin{equation}
    \textbf{y}_{p} = \boldsymbol{\Phi}\textbf{g} + \textbf{w}_p.
\end{equation}
Since \begin{math} {m \ll N} \end{math}, the vector \textbf{g} is a sparse vector with $m$ non zero blocks. Therefore, the received signal $\textbf{y}_{p}$ can be expressed as a linear combination of $m$ submatrices perturbed by the noise. Furthermore, the BS knows the pilot symbols and the SCMA codebook. Therefore, $\boldsymbol{\Phi}$ is available at the BS. The main task of the BS is to identify the $m$ submatrices in the received vector. For example, if the first and the third devices are active, then $\boldsymbol{\phi}_1$ and $\boldsymbol{\phi}_3$ are the components of $\textbf{y}_{p}$. Hence, the AUD problem can be formulated as the support identification problem: 
\begin{equation}
    \hat\Omega = \arg\min_{\mid\Omega\mid=m} \frac{1}{2} \parallel \textbf{y}_p- \Phi_\Omega\textbf{g}_\Omega \parallel_2^2.
\end{equation}

\section{Deep Neural Network Based AUD}
\subsection{Proposed Solution Approach}
\begin{figure}[t]
    \center
    \includegraphics[width=\linewidth]{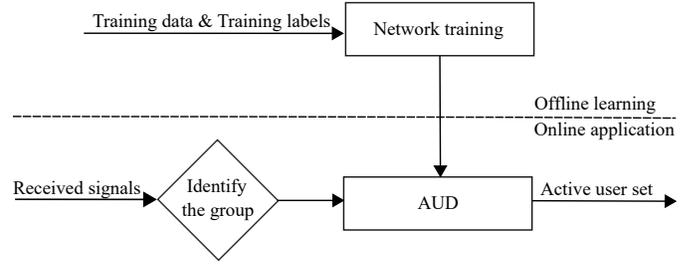}
    \caption{Block diagram of the proposed active user detection architecture.}
    \label{fig:Proposed_model}
\end{figure}

This study aims to pave the way for the practical implementation of how mMTC sparse signal processing technologies enable accurate and efficient AUD under the grant-free access scheme. However, the key characteristic of the mMTC traffic is that the device activity pattern is typically sporadic and therefore at any given time frame, only a small fraction of the massive devices are active. By exploiting this fact, the AUD problem can be formulated as a group based identification and deep learning (DL)-based multi-label classification. The block diagram of the proposed architecture is given in Fig.~\ref{fig:Proposed_model}. The BS first gets the received signals and identifies the group. It then detects the active devices by using the pre-trained AUD model. 

We propose to cluster the users into different groups using an appropriate clustering algorithm. For simplicity, we assume that only one group is active at any given time, though multiple devices within the same group can be active at the same time. This allows us to reuse the same SCMA codebook within different groups, which increases the utilization of the codebook and creates the possibility to design and use a small set of codebooks for the mMTC scenario. 

We develop two DL-based algorithms to identify the active devices inside any given group. There are no group-specific factors considered in the DL-based design. Therefore, it can be used for the non-group scenario as well. The specific objective of DL-based AUD is to identify the nonzero entries of $\textbf{g}$, not to recover the values in those entries. We design the DL-based algorithm with the objective of learning the complicated nonlinear mapping between the received signal $\textbf{y}_{p}$ and the support of $\textbf{g}$. Subsequently, the AUD can be modeled as
\begin{equation}
    \hat\Omega = g(\textbf{y}_{p};\Theta),
\end{equation}
where $g$ is the nonlinear mapping between the received signal and support of $\textbf{g}$, and $\Theta$ is the set of weights and biases of the neural network. 
\subsection{Deep Neural Network-AUD Structure}

The specific objective of this DL-based architecture is to find out $g$ characterized by $\Theta$ given $\textbf{y}_{p}$ closest to the optimal mapping function $g^*$. We exploit two different architectures to solve this problem. These are deep neural network-AUD and ResNet-AUD. Both have multiple building blocks of layers and an intrinsic dimensionality structure which learns the complicated correlation between the input signal and the support. The two DL-AUD architectures consist of several hidden layers; fully-connected (FC) layers, rectified linear unit (ReLU) layer, dropout layer, and sigmoid layer with the batch normalization. In proposed models, the input is the received signal at the BS, and the output is the labels of active MTDs. In this study, we generate the set $\mathcal{Q}$ of $Q$ training data $(\textbf{y}_{p}^{(1)},\dots,\textbf{y}_{p}^{(Q)})$ for each training iteration. Here, $\textbf{y}_{p}^{(q)}$ is a complex vector. Therefore we split real and imaginary parts separately and stack them as a vector input to the system given by 
\begin{equation}
    \textbf{y}_{p}^{(q)} = [\Re(y_{p,1}^{(q)}) \dots \Re(y_{p,L}^{(q)}) \Im(y_{p,1}^{(q)}) \dots \Im(y_{p,L}^{(q)})]^T.
\end{equation}
Details of the implementation of each AUD scheme are given in the following subsections.

\subsubsection{ResNet-AUD}
\begin{figure*}[tb]
    \center
    \includegraphics[width=\linewidth]{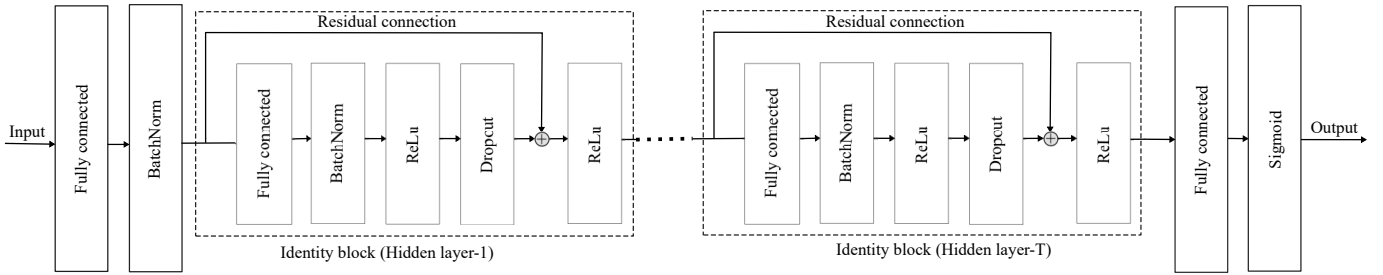}
    \caption{The detailed structure of the proposed ResNet based AUD.}
    \label{fig:ResNet}
\end{figure*}

Fig.~\ref{fig:ResNet} shows the detailed structure of the proposed ResNet based AUD. First, the input data passes through the FC layer and the output vector \begin{math}\textbf{z}^{(q)} \in \mathbb{R}^{\alpha \times 1}\end{math}of the FC layer is given by
\begin{equation}
    \textbf{z}^{(q)} = \textbf{W}^{in}\textbf{y}_{p}^{(q)} + \textbf{b}^{in}, \ \text{for} \ q = 1, \dots, Q,
\end{equation}
where \begin{math}\textbf{W}^{in} \in \mathbb{R}^{\alpha \times 2L} \end{math} is the initial weight, and \begin{math}\textbf{b}^{in} \in \mathbb{R}^{\alpha \times 1}\end{math} is the initial bias of the model. The next layer is batch normalization where $Q$ output vectors are stacked in mini-batch form and normalized to have zero mean and unit variance. The output $\Tilde{\textbf{z}}^{(q)}$ of the batch normalization is given by 
\begin{equation}
    \Tilde{z}_i^{(q)} = \beta\bigg(\frac{z_i^{(q)}-\mu_{B,i}}{\sqrt{\sigma_{B,i}^2}}\bigg) + \gamma,\ \text{for} \ i = 1, \dots, \alpha,
\end{equation}
where $\alpha$ is the hyper-parameter representing the width of the hidden layers, $\beta$ is the scaling parameter, $\gamma$ is the shifting parameter, $\mu_{B,i}$ is the batch-wise mean, and $\sigma_{B,i}^2$ is the batch-wise variance. At this layer, the architecture controls the variation of the wireless channel and noise levels to extract the internal features. After the batch normalization layer, the output vector passes through several identical blocks. Here, each block consists of a FC layer, batch normalization , ReLu activation unit, and a dropout layer. Then each identical block connects with a residual connection and a ReLu activation unit. The output of the FC and batch normalization layers inside each identical block is given in (9) and (10) with the corresponding layer parameters. Then, a nonlinear activation function is applied to the output of the batch normalization layer to determine whether the information passing through the hidden unit is activated. The output of the general ReLu function is given by $f(x) = max(x,0)$. After that, a dropout layer is added to regularize the model to avoid the risk of overfitting and to distinguish the correct support. The dropout layer is designed as a Bernoulli random variable $Bern(P_{dr})$ with a dropout probability of $P_{dr}$. During the training, a number of activated hidden layer outputs are randomly dropped out by removing the connectivity of incoming and outgoing layers of the dropped units. In effect, each update to a layer during training is performed with a different perspective of the configured layer. Therefore, the ambiguity of the activation patterns among correlated supports can be improved, which reduces the generalization error in the deep neural network. 
After having gone through all identical hidden layers, the last FC layer produces $N$ output values, which is equal to the total number of MTDs. The output vector is given by
\begin{equation}
    \textbf{z}^{out} = \textbf{W}^{out}\bigg( \Tilde{\textbf{z}} + \sum_{t=1}^T \Bar{\textbf{z}}^{(t)} \bigg) + \textbf{b}^{out},
\end{equation}
where $\Bar{\textbf{z}}^{(t)}$ is the output of the layer t, \begin{math}\textbf{W}^{out} \in \mathbb{R}^{N \times \alpha} \end{math}, and \begin{math}\textbf{b}^{out} \in \mathbb{R}^{N\times 1} \end{math} are the corresponding weight and bias, respectively. Then, sigmoid layer ($sigmoid(x) = \frac{1}{1+e^{-x}}$) consider each output values separately and produce $N$ independent probability values $(\hat{p}_1, \dots , \hat{p}_N)$, which represent the likelihood of the devices being active. Finally, the true support of the architecture chosen from the sigmoid outputs. 

\subsubsection{Deep Feed Forward Neural Network-AUD}
The proposed architecture consists of several FC layers with ReLu activation unit, and dropout layer. In particular this AUD model is simpler and has fewer layers than ResNet model. The output of each FC layer is the same as given in (9) with a different initial weight \begin{math}\textbf{W}^{in} \in \mathbb{R}^{\alpha \times 2L} \end{math}, and bias \begin{math}\textbf{b}^{in} \in \mathbb{R}^{\alpha \times 1}\end{math}. The ReLu and dropout functions are the same as in the previous model with different dropout probabilities. After passing through several hidden layers, the output of the network is given by 
\begin{equation}
    \textbf{z}^{out} = \textbf{W}^{out}\textbf{y}_{p}^{(q)} + \textbf{b}^{out},
\end{equation}
where \begin{math}\textbf{W}^{out} \in \mathbb{R}^{N \times \alpha} \end{math}, and \begin{math}\textbf{b}^{out} \in \mathbb{R}^{N\times 1} \end{math} are the corresponding weight and bias, respectively. Then sigmoid layer maps output values into $N$ independent probabilities and eventually the estimate of the support is obtained by sigmoid outputs. 

\subsection{Deep Neural Network-AUD Training Data Generation}
The approach taken here is classification. In particular, the classiﬁcation learning algorithm takes a collection of labeled examples as inputs and produces a model that can take an unlabeled data as input and can directly give a label as output. In neural network training, collecting a massive amount of real data and converting that data into inputs which the network can work with are the challenging aspects. However, in mMTC network system, we may use synthetically generated data for training, validation, and testing. 

In particular, we can say the AUD process is essentially the same as the support identification, and all channel components are accommodated in the input sparse vector \textbf{g}, not the system matrix $\boldsymbol{\Phi}$. Hence, the DL-AUD architecture only needs to learn the codebook matrix, which is known as a priori to the BS. Thus to produce data, we first generate a random noise vector \textbf{w} and a random block sparse vector \textbf{g} using the SCMA codebook, whose support is used as the label, and then generate $\textbf{y}_{p}$ by applying (2). Furthermore, in order to ensure the realistic nature of the data, we do not carry out any reduction of randomness~\cite{8761407}.

The amount of training data depends on the number of devices and the number of active devices. In our network, with $N$ MTDs in total, there are $\binom{N}{m}$ different combinations of labels for $m$ active MTDs. The number of training samples increases quickly as $m$ grows. Therefore, instead of generating the training data with a random active user number, we fix the number of active MTDs. This assumption is reasonable in the mMTC network because the number of active devices remains unchanged for several frames or time slots, and the number of active devices is much smaller than the total number of devices in the network~\cite{8444464}. However, we do not assume the number of active devices is known as a priori. Massive training data can be generated in this way, and the training operation of DL-AUD can be performed offline. Once the internal parameters of DL-AUD are learned, we can apply them to the actual wireless environment. In realistic scenarios, the network can be re-trained with long periodicity because the machine learning models learn the system conditions. Further, re-training is unnecessary till if there are no drastic changes in the environment and MTDs.

\section{Simulations Results}
\subsection{Simulation Setup}
In this section, we explore the performance of the two proposed deep neural network-AUD schemes with no prior CSI. We set the total number of MTDs inside a group to $6$ ($N$ = $6$) and we use $4$ subcarriers ($L$ = $4$) in each transmission. The overloading factor of the SCMA codebook is set to $150\%$. This simulation setup can be extended to any number of MTDs depending on mMTC requirement. Each device transmits pilot symbols using the quadrature phase shift keying (QPSK). For the channel model, we use an independent Rayleigh fading coefficient for each device. 

In both deep neural network-AUDs, we generate $10^6$ samples for training, validation, and testing. We set the learning rate to $0.001$, batch size to $1000$, and epoch to $20$, $T = 9$ is for ResNet-AUD, and the number of layers in DFF-AUD is 12. We use binary cross entropy as the loss function, Adam optimizer, ReLu activation function in hidden layers, and dropout probability $P_{dr}$ = $0.1$. Moreover, we use Keras with Tensorflow backend for the training, validation, and testing of the proposed two deep neural network-AUD schemes. As a performance metric during the training and validation period, we use confusion matrix that reports the number of false positives $(\textbf{fp})$, false negatives $(\textbf{fn})$, true positives $(\textbf{tp})$, and true negatives $(\textbf{tn})$. Furthermore, we examine precision $(\frac{\textbf{tp}}{\textbf{tp + fp}}$), recall ($\frac{\textbf{tp}}{\textbf{tp + fn}})$, and area under the receiver operating characteristic curve (AUC) during the hyperparameter tuning process to develop a robust model. We use the probability of detection (Recall) $(P_D)$ = $(\frac{\textbf{tp}}{\textbf{tp + fn}})$ and probability of misdetection $(P_M)$ = $(\frac{\textbf{fn}}{\textbf{tp + fn}})$ to show the performances. In addition, we evaluate the positive predictive values (precision) to show the correctness of the detected devices. Furthermore, we examine the performances of conventional least squares-BOMP (LS-BOMP)~\cite{8570860} and complex approximate message passing (C-AMP) algorithm~\cite{6478821} for comparison purposes. LS-BOMP minimizes the least-squares of a linear system with ~$\ell_0$ constraint and estimate the active devices with the knowledge of sparsity. C-AMP uses state evolution to estimates the active devices without the knowledge of sparsity. Also, we examine the performances of both LS-BOMP and C-AMP without CSI. 


\subsection{Simulation Results}
\begin{figure}[ht]
    \begin{center}
        \includegraphics[width=\linewidth]{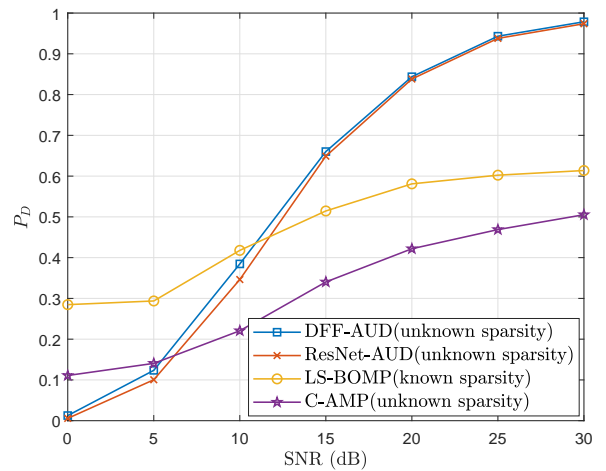}
    \end{center}
    \caption{Probability of detection versus the SNR for three different AUD schemes $(m = 1)$.}
    \label{fig:1AU}
\end{figure}

In Fig.~\ref{fig:1AU}, we investigate the probability of detection of proposed deep neural network-AUD schemes. We plot the probability of detection versus SNR for $m=1$. We observe that both deep neural network-AUD schemes outperform the conventional LS-BOMP from SNR = $10$ and C-AMP from SNR = $5$. Furthermore, the proposed deep feed forward neural network-AUD (DFF-AUD) outperforms all other schemes. Clearly, we can observe that DFF-AUD achieves around $20\%$ gain over LS-BOMP and $40\%$ gain over C-AMP at SNR = $20$ dB.

\begin{figure}[ht]
    \begin{center}
        \includegraphics[width=\linewidth]{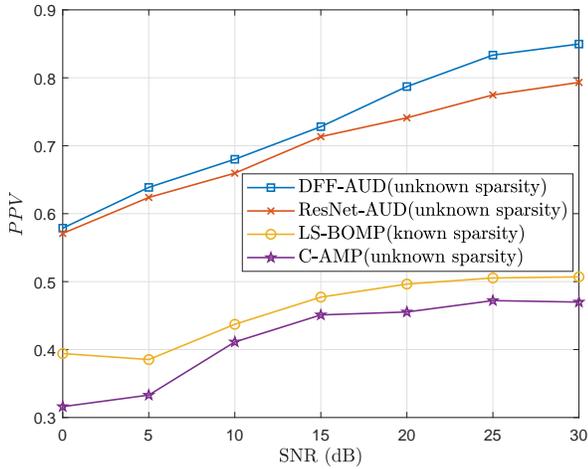}
    \end{center}
    \caption{Positive predictive values versus the SNR for three different AUD schemes $(m = 1)$.}
    \label{fig:1AU_PPV}
\end{figure}

In Fig.~\ref{fig:1AU_PPV}, we show the positive predictive value versus SNR for $m=1$.
This plot clearly shows that how many identified active devices are truly functional. Therefore, combine with Fig.~\ref{fig:1AU} we perceive the trustworthiness of the deep neural network-AUD schemes compare to the LS-BOMP and C-AMP approaches. 
Furthermore, considering both probabilities of detection and positive predictive value for the AUD problem provides the approach to control the type-I and type-II errors. 

\begin{figure}[th]
    \begin{center}
        \includegraphics[width=\linewidth]{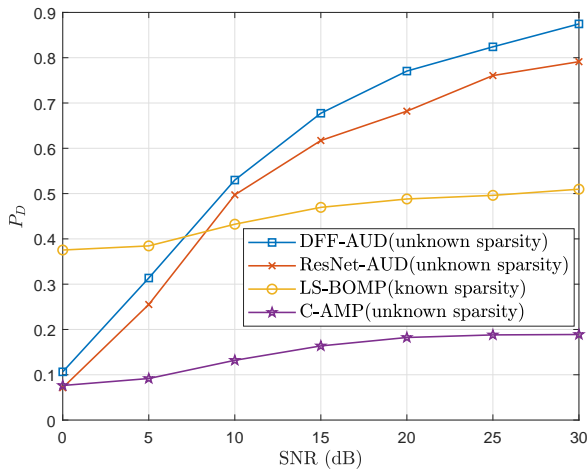}
    \end{center}
    \caption{Probability of detection versus the SNR for three different AUD schemes $(m = 2)$.}
    \label{fig:2AU-PoD}
\end{figure}

In Fig.~\ref{fig:2AU-PoD}, we evaluate the probability of detection of proposed deep neural network-AUD schemes for $m = 2$. We can clearly see the DFF-AUD outperforms the conventional LS-BOMP and C-AMP schemes. For example, DFF-AUD achieves twice the probability of detection at SNR = $30$ dB. It can be observed that the performance of the conventional schemes is not appealing when the number of active devices is high because of the increased correlation of the system matrix. 

\begin{figure}[ht]
    \begin{center}
        \includegraphics[width=\linewidth]{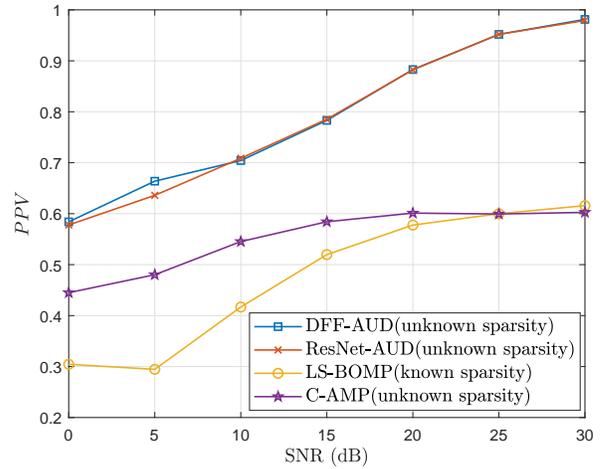}
    \end{center}
    \caption{Positive predictive values versus the SNR for three different AUD schemes $(m = 2)$.}
    \label{fig:2AU_PPV}
\end{figure}

In Fig.~\ref{fig:2AU_PPV}, we investigate the positive predictive values versus SNR for $m=2$. We observe the reliability of the deep neural network-AUD schemes compare to the LS-BOMP and C-AMP approaches. Furthermore, with the results of Fig.~\ref{fig:2AU-PoD}, we can observe better harmonic mean (known as $F1$-score) of the system $(\frac{PPV \times P_D}{PPV + P_D})$ than the conventional approaches. Furthermore, the $F1$-score collectively measures how many detected devices are truly active and how many active devices are detected among the total active devices.

\begin{figure}[ht]
    \begin{center}
        \includegraphics[width=\linewidth]{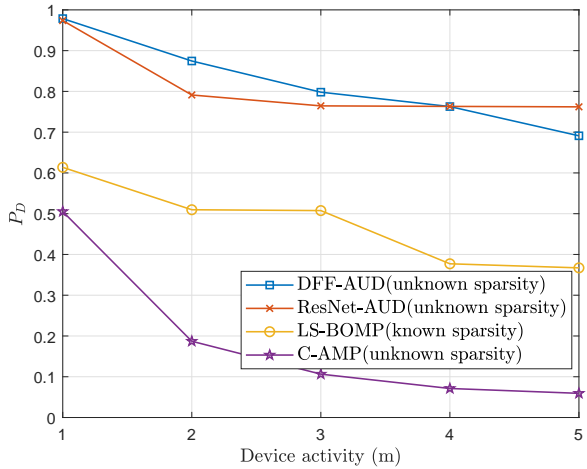}
    \end{center}
    \caption{Probability of detection versus number of active devices (SNR = $30$ dB).}
    \label{fig:All}
\end{figure}

In Fig.~\ref{fig:All}, we plot probability of detection versus active devices at SNR = $30$ dB. We observe that both deep neural network-AUD schemes outperform conventional LS-BOMP and C-AMP schemes across the entire region. When the number of active devices is $3$ $(m = 3)$ at SNR = $30$ dB, $P_D$ of DFF-AUD is $80\%$ while LS-BOMP is $50\%$ and C-AMP is $10\%$. Furthermore, we observe that the probability of detection of LS-BOMP is very small for active devices above $4$. Clearly, we can see the robustness of the deep neural network-AUD schemes with an increased number of active devices. In accordance with the present results, when $m$ increases, the correlation associated with the active devices increases, which leads to performance degradation. However, the deep neural network learns the system matrix during the training phase, while the conventional scheme does not. In addition, ResNet-AUD performs better than DFF-AUD for higher number of active devices. 
\begin{figure}[ht]
    \begin{center}
        \includegraphics[width=\linewidth]{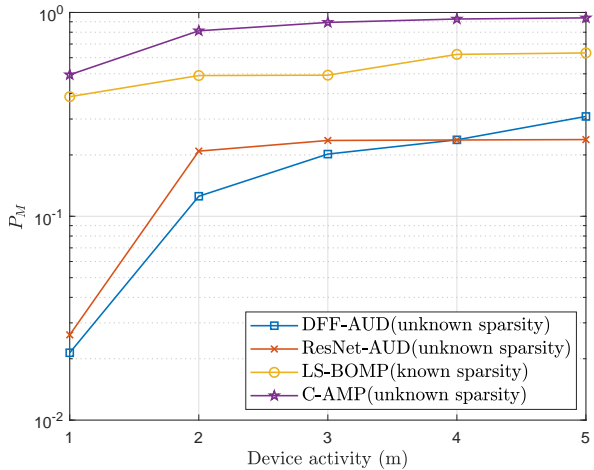}
    \end{center}
    \caption{Probability of misdetection versus number of active devices (SNR = $30$ dB).}
    \label{fig:Missdet}
\end{figure}

In Fig.~\ref{fig:Missdet}, we investigate the probability of misdetection for various number of active devices. We can clearly see that DFF-AUD outperforms the conventional AUD approaches by a large margin. For example, there will be $21$ misdetections $(P_M = 0.021)$ out of $10^3$ samples in DFF-AUD $(m = 1)$. 

\section{Conclusion}
In this paper, we have proposed two novel group-based deep neural network AUD schemes for the grant-free SCMA system in mMTC uplink framework. We have trained both schemes by feeding the training data to learn the nonlinear mapping between the received signal and the activity vector using an appropriate loss function. We have performed extensive numerical simulations to validate the performance of the proposed schemes. Simulation results have shown that they outperform the conventional LS-BOMP algorithm and C-AMP by a notable margin. 
These findings have significant implications for detecting active users without the knowledge of CSI and prior knowledge of the device sparsity level and maximizing the utilization of the SCMA codebook.

\bibliographystyle{IEEEtran}
\bibliography{main.bib}

\end{document}